\documentclass[12pt]{article}
\usepackage{amssymb}
\usepackage{epsfig}

\catcode`\@=11
\def\numberbysection{\@addtoreset{equation}{section}
        \def\theequation{\thesection.\arabic{equation}}}

\def\beq{\begin{equation}}
\def\eeq{\end{equation}}
\numberbysection
\begin{document}
\begin{titlepage}
\begin{center}
\hfill DFF  2/9/02 \\
\vskip 1.in {\Large \bf Noncommutative instantons on $d=2n$ planes
from matrix models} \vskip 0.5in P. Valtancoli
\\[.2in]
{\em Dipartimento di Fisica, Polo Scientifico Universit\'a di Firenze \\
and INFN, Sezione di Firenze (Italy)\\
Via G. Sansone 1, 50019 Sesto Fiorentino, Italy}
\end{center}
\vskip .5in
\begin{abstract}
In the case of an invertible coordinate commutator matrix
$\theta_{ij}$, we derive a general instanton solution of the
noncommutative gauge theories on $d=2n$ planes given in terms of
$n$ oscillators.
\end{abstract}
\medskip
\end{titlepage}
\pagenumbering{arabic}
\section{Introduction}

The concept of noncommutative geometry \cite{14}-\cite{13} has
nowadays acquired a central role in the study of possible
extensions of gauge theories. Plausible reasonings based on
quantum mechanics and general relativity imply that it is
necessary, at least at Planck scale, to replace space-time
coordinates by some noncommutative structure \cite{29}-\cite{30}.

Many non perturbative aspects of noncommutative gauge theories
have recently been explored; between them we can recall the Morita
equivalence of the noncommutative torus, and the non perturbative
structure of noncommutative gauge theories,i.e. the instantons (
in Euclidean space-time )
\cite{6}-\cite{8}-\cite{11}-\cite{20}-\cite{21}-\cite{22}.

The first examples of noncommutative instantons were given by
Nekrasov and Schwarz \cite{11} who modified the
Atiyah-Drinfeld-Hitchin-Manin (ADHM) construction \cite{18} to the
noncommutative case, and showed that on noncommutative $R^4$ non
singular instantons exist even for the $U(1)$ gauge group.

It is interesting to observe that introducing noncommutativity -
which from the point of view of the star product corresponds to
adding a complicated set of higher derivative interactions - can
in fact greatly simplify the construction of soliton solutions.

Our point of view is searching a new systematic method of
classifying soliton solutions starting from the Matrix model
approach, instead of the twistor approach of Nekrasov and Schwarz
\cite{23}-\cite{25}-\cite{3}-\cite{10}.

It is well known that matrix model, suitable defined, encodes the
noncommutative gauge theory on a $d=2n$ plane, by expanding the
matrices $X_i$ as a sum of a background ( representing the
noncommutative coordinate system ) and fluctuations, which
represent the gauge connection. In this paper we explore, by using
different matrix models, i.e. the unitary matrix model and the
hermitian one, the characterization of soliton solutions in terms
of constraints in the matrices $X_i$, and solve them with an easy
method.

The commonly accepted definition of soliton
\cite{1}-\cite{2}-\cite{4}-\cite{5}-\cite{7}-\cite{9}-\cite{12}-\cite{15}-\cite{16}-\cite{17}-\cite{19}-\cite{24}
is by the configuration

\beq U^{\dagger} X^i U \label{11} \eeq

obtained with a quasi-unitary operator $U$ that satisfies

\beq U U^{\dagger} = 1 \ \ \ \ \ \ \  U^{\dagger} U = 1 - P_0
\label{12} \eeq

where $P_0$ is a finite projection operator.

 In this article, instead of searching the general solution of this operator $U$
directly in $2n$-dimensions, we apply the duality transformation
\cite{27}-\cite{28} that maps a $2n$-dimensional plane in two
dimensions, since then the quasi-unitary operator $U$ is known to
be simply the shift operator. Using back again the duality
transformation one can easily write down general formulas for the
quasi-unitary operator $U$ and instanton configuration $X_i$ in
every even dimension.

The solution is based on a representation in terms of
$n$-oscillators in $d=2n$ dimensions, whose Hilbert space is the
tensorial product of $n$ Hilbert spaces of one oscillator.

The duality transformation is based on the isomorphism between $n$
quantum numbers and only one quantum number, which allows us to
reduce the general solution in $d=2n$ dim. to a solution with a
single oscillator, i.e. in $d=2$ dimensions.

\section{Matrix models and noncommutative gauge theories}

In this paper we study two types of matrix models, i.e. the
unitary matrix model and the hermitian matrix model, and their
relation with noncommutative gauge theory in $d=2n$ dimensions.

Let us start with $d=2$ and the unitary matrix model

\beq S = \frac{\beta_0}{\theta_0^2} Tr [ ( U_1 U_2 -
e^{-i\theta_0} U_2 U_1 ) ( h.c.) ] \ \ \ \ \ \ \ \theta_0 =
\frac{2\pi}{N} \label{21} \eeq

where $U_1, U_2$ are two independent unitary matrices $N \times N
$. In order to define a noncommutative $U(1)$ gauge theory in two
dimensions on a noncommutative plane, a certain $N\rightarrow
\infty$ limit must be taken.

The model has a positive definite action $S$, with its minimum
reached at

\beq U_1^{(0)} U_2^{(0)} = e^{-i\theta_0} U_2^{(0)} U_1^{(0)}
\label{22} \eeq

which are the commutation relations of a noncommutative torus. In
order to define the fluctuation around this background solution we
define:

\beq U_1 = U_1^{(0)} e^{i \sqrt{\theta \theta_0} a_1 ( U_1^{(0)},
U_2^{(0)} )} \ \ \ \ \ \ \ U_2 = U_2^{(0)} e^{i \sqrt{\theta
\theta_0} a_2 ( U_1^{(0)}, U_2^{(0)} )} \label{23} \eeq

where $a_1$ and $a_2$ are two hermitian matrices, that can be
developed, as in a Fourier series, in terms of the two basic
matrices $U_1^{(0)}, U_2^{(0)}$. By using simply the commutation
relations, one finds:

\begin{eqnarray} U_1 U_2 & = & U_1^{(0)} U_2^{(0)} e^{i \sqrt{\theta \theta_0}
a_1( U_1^{(0)} e^{-i\theta_0} , U_2^{(0)} )} e^{i \sqrt{\theta
\theta_0} a_2( U_1^{(0)}, U_2^{(0)} )} \nonumber \\
U_2 U_1 & = & U_2^{(0)} U_1^{(0)} e^{i \sqrt{\theta \theta_0} a_1(
U_1^{(0)}, U_2^{(0)} e^{i\theta_0} )} e^{i \sqrt{\theta \theta_0}
a_2( U_1^{(0)}, U_2^{(0)} )} . \label{24}
\end{eqnarray}

Now we consider the $N\rightarrow \infty$ limit ( i.e. $\theta_0
\rightarrow 0$ ). In this case it is possible to define two
noncommutative coordinates from the commutation relations
(\ref{22})

\beq U^{(0)}_1 = e^{i \sqrt{\frac{\theta_0}{\theta}} \hat{x}_2} \
\ \ \ \ U^{(0)}_2 = e^{-i \sqrt{\frac{\theta_0}{\theta}}
\hat{x}_1} \ \ \ \ \ \Rightarrow \ \ \ [ \hat{x}_1, \hat{x}_2 ] =
i \theta  . \label{25} \eeq

In terms of them, the fluctuations can be recast as follows

\begin{eqnarray}
U_1 U_2 & = & U^{(0)}_1 U^{(0)}_2 e^{i \sqrt{\theta \theta_0} a_1
( \hat{x}_1, \hat{x}_2 - \sqrt{\theta \theta_0} ) } e^{i
\sqrt{\theta \theta_0} a_2 ( \hat{x}_1, \hat{x}_2 ) } \nonumber \\
U_2 U_1 & = & U^{(0)}_2 U^{(0)}_1 e^{i \sqrt{\theta \theta_0} a_1
( \hat{x}_1 - \sqrt{\theta \theta_0}, \hat{x}_2 ) } e^{i
\sqrt{\theta \theta_0} a_2 ( \hat{x}_1, \hat{x}_2 ) } . \label{26}
\end{eqnarray}

By expanding the exponentials and keeping only the relevant terms,
one finds

\begin{eqnarray} U_1 U_2 - e^{-i\theta_0} U_2 U_1 & = & U_1^{(0)}
U_2^{(0)} [ i \sqrt{\theta \theta_0} ( a_1 ( \hat{x}_1, \hat{x}_2
- \sqrt{\theta \theta_0} ) - a_1 ( \hat{x}_1 , \hat{x}_2 ) ) \nonumber \\
& - & \frac{\theta\theta_0}{2} ( a^2_1 + a^2_2 + 2 a_1 a_2 ) - i
\sqrt{\theta \theta_0} ( a_2 ( \hat{x}_1 - \sqrt{\theta\theta_0},
\hat{x}_2 ) - a_2 ( \hat{x}_1, \hat{x}_2 )) \nonumber \\
& + & \frac{\theta \theta_0}{2} ( a^2_1 + a^2_2 + 2 a_2 a_1 ) ]
+ O ( \sqrt{\theta^3_0}) = \nonumber \\
& = & U^{(0)}_1 U^{(0)}_2 ( - i \theta \theta_0 ( \partial_2 a_1 -
\partial_1 a_2 ) - \theta \theta_0 [a_1, a_2]) \nonumber \\
& = & i \theta \theta_0 U^{(0)}_1 U^{(0)}_2 ( \partial_1 a _2 -
\partial_2 a_1 + i [ a_1, a_2 ] ) = i \theta \theta_0 U^{(0)}_1
U^{(0)}_2 F_{12} + O(\sqrt{\theta^3_0}) . \nonumber \\
& \ & \label{27}
\end{eqnarray}

Strictly speaking, defining a derivative with respect to a
noncommutative coordinate is not properly defined. One should pass
from operators to their symbols, i.e. commutative functions , as
in \cite{27} , and introduce, instead of the products of operators
the star product of their corresponding symbols; however we will
never need to use the symbols and we will omit them in the
following.

The action which survives in the $N \rightarrow \infty$ limit is
the $U(1)$ noncommutative Yang-Mills action:

\beq S = \lim_{\theta_0 \rightarrow 0 } \frac{\beta_0}{\theta_0^2}
Tr ( U_1 U_2 - e^{-i\theta_0} U_2 U_1 ) ( h.c. ) = \beta_0
\theta^2 Tr F_{12}^2 . \label{28} \eeq

We have reached the noncommutative plane from a careful
parametrization of the commutation relation (\ref{22}), which
represents instead a noncommutative torus as a background.

Note the importance of the presence of the scaling factor $N^2$ in
front of the action $S$, to reach a finite action in the $N
\rightarrow \infty$ limit.

It is not difficult to relate this model to a hermitian matrix
model directly. The two independent unitary matrices $U_1, U_2$
can be parameterized in terms of two infinite-dimensional
hermitian matrices as follows

\beq U_1 = e^{i \sqrt{\theta_0 \theta} X_1 } \ \ \ \ U_2 = e^{i
\sqrt{\theta_0 \theta} X_2 } . \label{29} \eeq

Then the commutation relation

\beq U_1 U_2 - e^{-i\theta_0} U_2 U_1 = - \theta \theta_0 ( [ X_1,
X_2 ] - \frac{i}{\theta} ) + O ( \sqrt{\theta^3_0}) \label{210}
\eeq

is equivalent to a commutation relation between the two
infinite-dimensional hermitian matrices.

Therefore, in the limit $\theta_0 \rightarrow 0$, the action $S$
becomes:

\beq S = - \beta_0 \theta^2 Tr ({ [ X_1, X_2 ] - \frac{i}{\theta}
)}^2 \label{211} \eeq

which again is positive definite and reaches its minimum on the
background

\beq [ X_1, X_2 ] = \frac{i}{\theta} \label{212} \eeq

which is satisfied by a noncommutative plane, with the position

\beq X_1 = \frac{\hat{x}_2}{\theta} \ \ \ \ X_2 = -
\frac{\hat{x}_1}{\theta} \ \ \ \Rightarrow [ \hat{x}_1, \hat{x}_2
] = i \theta . \label{213} \eeq

The corresponding equations of motion are identical to those of
the standard hermitian matrix model given by :

\beq S_0 = - \beta_0 \theta^2 Tr {[X_1 , X_2 ]}^2 \label{214} \eeq

i.e.

\beq [ X_j, [ X_i, X_j ]] = 0 \ \ \ \ \ \ i, j = 1,2 . \label{215}
\eeq

To find the direct connection between hermitian matrix model and
Yang-Mills action, it is enough to develop the hermitian matrices
as a background + fluctuations as follows :

\beq X_1 = \frac{\hat{x}_2}{\theta} + a_1 ( \hat{x}_1, \hat{x}_2 )
\ \ \ \ X_2 = - \frac{\hat{x}_1}{\theta} + a_2 ( \hat{x}_1,
\hat{x}_2 ) \label{216} \eeq

from which

\beq [ X_1, X_2 ] - \frac{i}{\theta} = - i F_{12} \label{217} \eeq

and the action becomes

\beq S = - \beta_0 \theta^2 Tr {( [ X_1, X_2 ] - \frac{i}{\theta}
)}^2 = \beta_0 \theta^2 Tr F_{12}^2 . \label{218} \eeq

\section{Instantons in two dimensions}

Let us recall the action for two-dimensional unitary matrix model

\beq S = \frac{\beta_0}{\theta^2_0} Tr [ ( U_1 U_2 -
e^{-i\theta_0} U_2 U_1 ) ( h.c.) ] \ \ \ \ \ \       \theta_0 =
\frac{2\pi}{N} . \label{31} \eeq

The search for instantons of the noncommutative planes is driven
by two criteria. The first is that an instanton is a solution of
the equations of motion of $S$, and the second is that this
solution must give a finite contribution to the action, also in
the $N \rightarrow \infty$ limit.

The classical equations of motion, obtained from a variation of
the basic elements of the theory , $U_1$ and $U_2$, are:

\beq [ U_i, V - V^{\dagger} ] = 0 \label{32} \eeq

where $ V = e^{-i\theta_0} U_1 U_2 U^{\dagger}_1 U^{\dagger}_2 $.

Between all the possible solutions to equation (\ref{32}), we find
that the instantons are characterized by the matrices of the form:

\beq \tilde{U}_1 = \left( \begin{array}{cc} U_1^{(0)} & 0 \\ 0 &
U_1^{(1)} \end{array} \right) \ \ \ \ \tilde{U}_2 = \left(
\begin{array}{cc} U_2^{(0)} & 0 \\ 0 & U_2^{(1)} \end{array}
\right) \label{33} \eeq

where $U_1^{(0)}$ and $U_2^{(0)}$ are two $(N-k) \times (N-k)$
unitary matrices satisfying

\beq U_1^{(0)} U_2^{(0)} = e^{\frac{2i\pi}{N-k}} U_2^{(0)}
U_1^{(0)} \label{34} \eeq

and $ U_1^{(1)}, U_2^{(1)} $ are two $k \times k$ diagonal unitary
matrices.

Computing the action (\ref{31}) on the solution (\ref{33}) is not
a difficult task:

\beq S = \frac{\beta_0}{\theta^2_0} [ 2 ( 1 - cos ( \frac{2\pi}{N}
- \frac{2\pi}{N-k} )) (N-k) + 2 ( 1 - cos \frac{2\pi}{N} ) k ] .
\label{35} \eeq

The first term is order $O(\frac{1}{N})$ and vanishes in the
$N\rightarrow \infty$ limit, while the second term is finite and
gives as contribution:

\beq S = \beta_0 k . \label{36} \eeq

Such class of solution has a counterpart in the hermitian matrix
model. Recall the action of the hermitian matrix model

\beq S = - \frac{1}{2 g'^{2}} Tr ( [ X_1, X_2] - \frac{i}{\theta}
)^2 . \label{37} \eeq

It is not difficult to extract from (\ref{33}) that in the $N
\rightarrow \infty$ limit :

\beq \lim_{\theta_0 \rightarrow 0 } \frac{1}{\theta_0} (
\tilde{U}_1 \tilde{U}_2 - e^{-i\theta_0} \tilde{U}_2 \tilde{U}_1 )
=  \lim_{\theta_0 \rightarrow 0} \left( \begin{array}{cc} 0 & 0 \\
0 &  \frac{ 1 - e^{-i \theta_0}}{\theta_0} \end{array} \right)
U^{(1)}_1 U^{(2)}_1 = i U^{(1)}_1 U^{(2)}_1 \left(
\begin{array}{cc} 0 & 0 \\ 0 & P_0 \end{array} \right) \label{38} \eeq

this commutation relation is equivalent to a $k \times k$
projector $P_0$, apart from an irrelevant phase.

Since

\beq \frac{1}{\theta_0} ( \tilde{U}_1 \tilde{U}_2 - e^{-i\theta_0}
\tilde{U}_2 \tilde{U}_1 ) = - \theta ( [X_1, X_2] -
\frac{i}{\theta} ) + O ( \sqrt{\theta_0^3}) \label{39} \eeq

we derive, that on an instanton solution, the following relation
holds:

\beq [ X_1, X_2 ] = \frac{i}{\theta} ( 1- P_0 ) . \label{310} \eeq

This condition will be the starting point of this article to
characterize an instanton solution on a noncommutative plane. The
instanton number is related to the rank $k$ of the projector
$P_0$, as it is clear from the evaluation of the solution
(\ref{33}) into the action :

\beq S = - \frac{2}{g'^2} ( \frac{i}{\theta} )^2 Tr P_0 =
\frac{1}{2 g'^2 \theta^2} k . \label{311} \eeq

Notice that, with a constant shift of the action (\ref{37}) with
respect to the usual hermitian matrix model, the evaluation of the
action is null on the vacuum solution and finite on the instanton
solution.

Notice also that both the action and the Chern class of $F_{12}$
are linear in the instanton number, and that the Yang-Mills
curvature $F_{12}$ is a projector:

\beq F_{12} = P_0 . \label{312} \eeq

The two dimensional plane is made by two coordinates which are
promoted to operators acting on a Hilbert space. The Hilbert space
is equivalent to the usual one oscillator Hilbert space, where the
coordinates $\hat{x}_1$, $\hat{x}_2$ are given in terms of
oscillator rising and lowering operators:

\begin{eqnarray} a & = & \sqrt{\frac{1}{2\theta}} ( \hat{x}_1 + i
\hat{x}_2 ) \ \ \ \ \ \ \ \ \ \overline{a} =
\sqrt{\frac{1}{2\theta}} ( \hat{x}_1 - i \hat{x}_2 ) \nonumber \\
a |n> & = & \sqrt{n} |n-1> \ \ \ \ \ \ \ \ \ \ \overline{a} |n> =
\sqrt{n+1} |n+1> . \label{313} \end{eqnarray}

Therefore the vacuum is characterized by the solution

\begin{eqnarray} X^1 + i X^2 & = &  - \frac{i}{\theta} ( \hat{x}_1 + i \hat{x}_2 )
= - i \sqrt{\frac{2}{\theta}} a = - i \sqrt{\frac{2}{\theta}}
\sum_{n=0}^{\infty} \sqrt{n+1} |n><n+1|  \nonumber \\
X^1 - i X^2 & = &  i \sqrt{\frac{2}{\theta}} \overline{a} =  i
\sqrt{\frac{2}{\theta}}  \sum_{n=0}^{\infty} \sqrt{n+1} |n+1><n| .
\label{314}
 \end{eqnarray}

Instead, in the Hilbert space, the operator that defines the
configuration of the $k$-instanton is given by :

\begin{eqnarray} X^1 + i X^2 & = & - i \sqrt{\frac{2}{\theta}} A_k
\ \ \ \ \ \ \ \ \ X^1 - i X^2  =  i \sqrt{\frac{2}{\theta}}
\overline{A}_k \nonumber \\
A_k & = & \sum_{n_1=0}^{k-1} \ c_{n_1} |n_1 ><n_1 | +
\sum_{n_1=k}^{\infty} \sqrt{n_1 +1 - k } |n_1><n_1+1| \nonumber \\
\overline{A}_k & = & \sum_{n_1=0}^{k-1} \ \overline{c}_{n_1} |n_1
><n_1 | + \sum_{n_1=k}^{\infty} \sqrt{n_1 +1 - k } |n_1+1><n_1|
\label{315}
\end{eqnarray}

where

\beq [ A_k, \overline{A}_k ] = 1 - P_0 . \label{316} \eeq

Till now we have discussed the case of the gauge group $U(1)$. It
is possible to describe with the same method the instanton
solution of $U(2)$ gauge group, or in general the $U(n)$ gauge
group, by enlarging the Hilbert space one-oscillator to the base

\beq |n; a > \ \ \ a = 0,1 \ \ \ {\rm for} \ U(2) \ \ \ {\rm or} \
\  a = 0,1,...,n-1 \ \ {\rm for } \ U(n) . \label{317} \eeq

Then a $k$-instanton solution of $U(2)$ can be characterized by
the complex lowering oscillator operator:

\beq \tilde{A}_k = \sum_{a=0}^1 ( \sum_{n_1=0}^{k-1} c_{n,a}
|n_1;a><n_1:a| + \sum_{n_1=k}^{\infty} \sqrt{n_1 +1 - k}
|n_1;a><n_1;a|) . \label{318} \eeq

Due to the duality relation between the $U(2)$ gauge group and
$U(1)$ gauge group on the two-dimensional noncommutative plane
\cite{27}-\cite{28}, it is possible to establish an isomorphism
between Hilbert spaces

\beq |n_1:a > = | 2 n_1 + a > . \label{319} \eeq

Therefore $\tilde{A}_k$, being a connection of $U(2)$ gauge group,
can be mapped to a $U(1)$ connection, due to (\ref{319}):

\begin{eqnarray} A_k^{dual} & = & \sum_{a=0}^1 (
\sum_{n_1=0}^{k-1} c_{2n_1+a} | 2n_1 + a>< 2n_1 + a| +
\sum_{n_1=k}^{\infty} \sqrt{n_1+1-k} |2n_1+a><2n_1+2+a| ) \nonumber \\
& = & \sum_{n_1=0}^{k-1} ( c_{2n_1} |2n_1 ><2n_1| + c_{2n_1+1}
|2n_1+1><2n_1+1| ) \nonumber \\
& + & \sum_{n_1=k}^{\infty} \sqrt{n_1+1-k} ( |2n_1><2n_1+2| +
|2n_1+1><2n_1+3|) . \label{320} \end{eqnarray}

Passing from $n_1 \rightarrow \frac{n_1}{2}$ we obtain

\begin{eqnarray} A_k^{dual} & = & \sum_{n_1=0}^{2k-1} c_{n_1}
|n_1><n_1| + \sum_{\frac{n_1}{2} = k}^{\infty} \sqrt{\frac{n_1}{2}
- k +1 }
( |n_1><n_1+2|  \nonumber \\
& + & |n_1+1><n_1+3|) . \label{321} \end{eqnarray}

While the second part is a gauge transformation of

\beq \sum_{n_1=2k}^{\infty} \ \sqrt{n_1-2k+1} |n_1><n_1 + 1|
\label{322} \eeq

and therefore equivalent to it, the first part produces into the
action an $U(1)$ instanton of type $2k$.

In general, by the same reasoning, a $U(n)$ $k$-instanton should
be equivalent to a $n \cdot k$ instanton on $U(1)$ on the
two-dimensional plane.

\section{Instantons on a four-dimensional noncommutative plane}

The generalization of the two-dimensional hermitian matrix model
(\ref{37}) on a four-dimensional noncommutative plane is given by
the equation

\beq S = - \frac{1}{4g^2} Tr ( [ X_i, X_j ] - i \Theta^{-1}_{ij}
)^2 \label{41} \eeq

where it is necessary to require that the coordinate commutator
matrix $\Theta_{ij}$ is invertible, i.e.

\beq [ \hat{x}_i , \hat{x}_j ] = i \Theta_{ij} \ \ \ \ \ det
\Theta_{ij} \neq 0 . \label{42} \eeq

Instead of using the commutation relations (\ref{42}) in full
generality, we now show how to diagonalize (\ref{42}) in a form
which is basically solved by two types of independent raising and
lowering oscillator operators.

Firstly we consider the commutation relations (\ref{42}),
rewritten in this form:

\begin{eqnarray} & \ & [ \hat{x}_1, \hat{x}_2 ]  =  i \tilde{\theta}_1 cos
\phi_1 \ \ \ \ \ [\hat{x}_1, \hat{x}_3 ] = i \tilde{\theta}_1 sin
\phi_1 cos \phi_2 \ \ \ \ \ [ \hat{x}_1, \hat{x}_4 ] = i
\tilde{\theta}_1 sin \phi_1 sin \phi_2 \nonumber \\
& \ & [ \hat{x}_2, \hat{x}_3 ]  =  i \theta_{23} \ \ \ \ \ [
\hat{x}_2, \hat{x}_4 ] = i \theta_{24} \ \ \ \ \ [ \hat{x}_3,
\hat{x}_4 ] = i \theta_{34} . \label{43} \end{eqnarray}

Let us define a new set of orthogonal cartesian coordinates :

\begin{eqnarray}
x'_1 & = & x_1 \nonumber \\
x'_2 & = & x_2 cos \phi_1 + x_3 sin \phi_1 cos \phi_2 + x_4 sin
\phi_1 sin \phi_2 \nonumber \\
x'_3 & = & x_3 sin \phi_2 - x_4 cos \phi_2 \nonumber \\
x'_4 & = & x_4 sin \phi_2 cos \phi_1 + x_3 cos \phi_2 cos \phi_1 -
x_2 sin \phi_1 . \label{44} \end{eqnarray}

Then the first part of the commutation relations (\ref{42})
reduces to:

\beq [ x'_1, x'_2 ] = i \tilde{\theta}_1 \ \ \ \ \ [ x'_1, x'_3 ]
= 0 \ \ \ \ \ [ x'_1, x'_4 ] = 0 . \label{45} \eeq

After the orthogonal transformation (\ref{44}), the other
commutation relations read

\beq [ x'_2, x'_3 ] = i \tilde{\theta}_2 sin \delta_1 sin \delta_2
\ \ \ \ \ [ x'_2, x'_4 ] = i \tilde{\theta}_2 sin \delta_1 cos
\delta_2 \ \ \ \ \ [x'_3, x'_4] = i \tilde{\theta}_2 cos \delta_1
. \label{46}  \eeq

Now we apply another orthogonal transformation:

\begin{eqnarray}
x''_1 & = & x'_1 \nonumber \\
x''_2 & = & x'_2 \nonumber \\
x''_3 & = & x'_3 sin \delta_2 + x'_4 cos \delta_2 \nonumber \\
x''_4 & = & x'_4 sin \delta_2 - x'_3 cos \delta_2 \label{47}
\end{eqnarray}

to reduce the commutation relations to the form:

\begin{eqnarray}
& \ & [ x''_2, x''_3 ] = i \tilde{\theta}_2 sin \delta_1 \nonumber
\nonumber \\
& \ & [ x''_2, x''_4 ] = 0 \nonumber \\
& \ & [ x''_3, x''_4 ] = i \tilde{\theta}_2 cos \delta_1 .
\label{48}
\end{eqnarray}

Finally, we apply a linear non-orthogonal transformation, which is
invertible if $\Theta_{ij}$ is a non degenerate matrix, given by :

\begin{eqnarray}
\tilde{x}_1 & = & x''_1 \nonumber \\
\tilde{x}_2 & = & x''_2 \nonumber \\
\tilde{x}_3 & = & cos \delta_1 x''_3 - sin \delta_1 x''_4 \nonumber \\
\tilde{x}_4 & = & x''_4 \label{49}
\end{eqnarray}

to find:
\begin{eqnarray} & \ & [ \hat{x}_1, \hat{x}_2 ]  =  i \theta_1
\ \ \ \ \ [\hat{x}_1, \hat{x}_3 ] = 0 \ \ \ \ \ [ \hat{x}_1,
\hat{x}_4 ] = 0  \nonumber \\
& \ & [ \hat{x}_2, \hat{x}_3 ]  =  0  \ \ \ \ \ [ \hat{x}_2,
\hat{x}_4 ] = 0  \ \ \ \ \ [ \hat{x}_3, \hat{x}_4 ] = i
\tilde{\theta}_2 cos^2 \delta_1 = i \theta_2 . \label{410}
\end{eqnarray}

This diagonal form of commutation relations will be used in the
following part of the section to define a general class of
four-dimensional instanton solutions.

To make contact with noncommutative Yang-Mills theory we decompose
the matrices $X_i$ as a sum of background plus fluctuations as
follows:

\begin{eqnarray}
X_1 & = & \frac{1}{\theta_1} \tilde{x}_2 + A_1 \ \ \ \ \ X_2 = -
\frac{1}{\theta_1} \tilde{x}_1 + A_2 \nonumber \\
X_3 & = & \frac{1}{\theta_2} \tilde{x}_4 + A_3 \ \ \ \ \ X_4 = -
\frac{1}{\theta_2} \tilde{x}_3 + A_4 \label{411}
\end{eqnarray}

and the action turns out to be

\beq S = \frac{1}{4 g'^2} Tr F_{ij}^2 \ \ \ \ \ F_{ij} =
\partial_i A_j - \partial_j A_i + i [ A_i, A_j ] . \label{412} \eeq

The vacuum is again characterized by the solution

\begin{eqnarray}
X^1 + i X^2 & = & - \frac{i}{\theta_1} ( x^1 + i x^2 ) = - i
\sqrt{\frac{2}{\theta_1}} a_1 = - i \sqrt{\frac{2}{\theta_1}} (
\sum_{n_1, n_2 =0}^{\infty} \sqrt{n_1+1} |n_1, n_2><n_1+1, n_2| )
\nonumber \\
X^3 + i X^4 & = & - \frac{i}{\theta_2} ( x^3 + i x^4 ) = - i
\sqrt{\frac{2}{\theta_2}} a_2 = - i \sqrt{\frac{2}{\theta_2}} (
\sum_{n_1, n_2 =0}^{\infty} \sqrt{n_2+1} |n_1, n_2><n_1, n_2+1| )
\nonumber \\
& \ & \label{413}
\end{eqnarray}

where

\beq [ a_1, \overline{a}_1 ] = [a_2, \overline{a}_2 ] = 1 \ \ \ \
[ a_1, a_2 ] = [ a_1, \overline{a}_2 ] = 0 . \label{414} \eeq

The definition, commonly accepted, of soliton is given by

\beq U^{\dagger} X^i U \label{415} \eeq

obtained by a quasi-unitary operator $U$ that satisfies

\beq U U^{\dagger} = 1 \ \ \ \ \ U^{\dagger} U = 1 - P_0
\label{416} \eeq

where $P_0$ is a finite projection operator.

A simple consequence of (\ref{416}) is that, for an instanton
solution we must study the following system of commutation rules

\beq [ X_i, X_j ] = i \Theta_{ij}^{-1} ( 1 - P_0 ) \label{417}
\eeq

and give the general and explicit solution of this system for a
generic finite projection operator $P_0$.

The instanton solution we look for is therefore given by a
deformation of the two raising and lowering operators $a_i ( i =
1,2)$ defined through

\begin{eqnarray}
X_1 + i X_2 & = & -i \sqrt{\frac{2}{\theta_1}} A_1 \nonumber \\
X_3 + i X_4 & = & -i \sqrt{\frac{2}{\theta_2}} A_2 \label{418}
\end{eqnarray}

and the postulated commutation relations for $A_i$ are:

\beq [ A_1, \overline{A}_1 ] = [ A_2, \overline{A}_2 ] = 1- P_0 \
\ \ \ \ [ A_1, A_2 ] = [ A_1, \overline{A}_2 ] = 0 . \label{419}
\eeq

Notice that, due to the duality relation between the tensorial
product of Hilbert spaces and the one-oscillator Hilbert space
\cite{27}-\cite{28}

\beq {\cal H} \times {\cal H} \rightarrow {\cal H} \label{439}
\eeq

based on the fact that a couple of numbers can be made isomorphic
to a number, for example

\beq ( n_1, n_2 ) \rightarrow \frac{ ( n_1+n_2 ) ( n_1+n_2+1 )}{2}
+ n_2 \label{440} \eeq

it is possible to relate the $U(1)$ four-dimensional connection
$(A_1, A_2)$ to a couple of two-dimensional connections of the
$U(1)$ theory in two dimensions.

Let us define two new quantum numbers :

\begin{eqnarray}
n & = & n_1 + n_2 \nonumber \\
k & = & n_2 \label{01}
\end{eqnarray}

and introduce a short notation for the state :

\beq | n_1, n_2 > = | \frac{n(n+1)}{2} + k > = | n ; k > .
\label{02} \eeq

A basis of the two-dimensional Hilbert space is determined by the
states

\beq | n ; k > \ \ \ \ \ 0 \leq k \leq n \ \ \ \ \forall n \in N .
\label{03} \eeq

In this paper however we will need to continue the notation
(\ref{02}) to states with $ k \geq n$, keeping in mind then the
equivalence relation

\beq | n ; k > = | n+1 ; k - n -1 > . \label{04} \eeq

In the two-dimensional basis, the generic finite projection
operator $P_0$ can be represented in the following form, up to an
isomorphism of the Hilbert space:

\beq P_0 = |0><0| + ... + | m-1 >< m-1 | \label{05} \eeq

which represents a configuration with instanton number $m$.

In the two-dimensional basis (\ref{03}) it is not difficult to
derive the quasi-unitary operator $U$ that produces the projection
operator $P_0$:

\begin{eqnarray} & \ & U U^{\dagger} = 1 \ \ \ \ \ \ U^{\dagger} U = 1 -
P_0 \nonumber \\
& \ & U = \sum_{n=0}^{\infty} \sum_{k=0}^n \ | n ; k >< n ; k + m
| \nonumber \\
& \ & U^{\dagger} = \sum_{n=0}^{\infty} \sum_{k=0}^n \ | n ; k + m
>< n ; k |. \label{06} \end{eqnarray}

In order to define the instanton configuration, we need to compute

\beq A^{(1)}_1 = U^{\dagger} a_1 U \ \ \ \ \ A^{(1)}_2 =
U^{\dagger} a_2 U \label{07} \eeq

and it is necessary to represent the lowering oscillator operator
$a_1, a_2$ in the new basis as follows:

\begin{eqnarray}
a_1 & = & \sum_{n=0}^{\infty} \sum_{k=0}^n \ \sqrt{ n + 1 - k } | n ; k >< n+1 ; k |
\nonumber \\
a_2 & = & \sum_{n=0}^{\infty} \sum_{k=0}^n \ \sqrt{ 1 + k } | n
 ; k >< n+1 ; k +1 |. \label{08} \end{eqnarray}

 Computing the products (\ref{07}) is then an easy task, and the
result is :

\begin{eqnarray}
A_1^{(1)} & = & \sum_{n=0}^{\infty} \sum_{k=0}^n \ \sqrt{ n + 1 -
k } | n ; k + m >< n+1 ; k + m |
\nonumber \\
A_2^{(1)} & = & \sum_{n=0}^{\infty} \sum_{k=0}^n \ \sqrt{ 1 + k }
| n ; k + m >< n+1 ; k + m + 1 |. \label{09} \end{eqnarray}

An obvious consequence of their definition is that they satisfy
the commutation rules (\ref{419})

\beq [ A_1^{(1)}, \overline{A}_1^{(1)} ] = [ A_2^{(1)},
\overline{A}_2^{(1)} ] = 1 - P_0 \ \ \ \ \ [ A_1^{(1)}, A_2^{(1)}
] = [ A_1^{(1)}, \overline{A}_2^{(1)} ] = 0 . \label{010} \eeq

For completeness we give the following formulas:

\begin{eqnarray}
& \ & A^{(1)}_1 A^{(1)}_2 = A^{(1)}_2 A^{(1)}_1 =
\sum_{n=0}^{\infty} \sum_{k=0}^n \sqrt{ ( n + 1 - k ) ( 1 + k ) }
| n ; k + m >< n + 2 ; k + m + 1 | \nonumber \\
& \ & A^{(1)}_1 \overline{A}^{(1)}_2 = \overline{A}^{(1)}_2
A^{(1)}_1 = \sum_{n=1}^{\infty} \sum_{k=1}^n \sqrt{ k ( n + 1 - k
) } | n ; k + m >< n  ; k + m - 1 | \nonumber \\
& \ & A^{(1)}_1 \overline{A}^{(1)}_1 - \overline{A}^{(1)}_1
A^{(1)}_1 = A^{(1)}_2 \overline{A}^{(1)}_2 - \overline{A}^{(1)}_2
A^{(1)}_2 = \sum_{n=0}^{\infty} \sum_{k=0}^n | n ; k + m >< n ; k
+ m | = 1 - P_0 .\nonumber \\
& \ &  \label{011} \end{eqnarray}

The equations of motion are also satisfied, since

\beq [ X_j , [ X_i, X_j ]] = 0 \ \ \ \Rightarrow \ \ \ [
A_i^{(1)}, P_0 ] = 0 \ \ \ i = 1,2 \label{012} \eeq

as one can immediately verify.

This is the basic solution. It is possible to add to the solution
(\ref{09}) an arbitrary configuration that, commuting with
everything, doesn't alter the commutation rules (\ref{419}) and
the equations of motion (\ref{012}) :

\begin{eqnarray}
& \ & A_1 = A^{(0)}_1 + A^{(1)}_1 \ \ \ \ \ \ \ A_2 = A^{(0)}_2 +
A^{(1)}_2 \nonumber \\
& \ & A_1^{(0)} = \sum_{i=0}^{m-1} \ c^1_i | i >< i | \ \ \ \ \ \
A_2^{(0)} = \sum_{i=0}^{m-1} \ c^2_i | i >< i | . \label{013}
\end{eqnarray}

In order to derive the configuration $A_1, A_2$ in terms of the
equivalent basis $|n_1, n_2>$, we need to pull back the duality
between two-dimensional and four-dimensional plane. The task is
cumbersome in the general case, and we will perform it only in the
simplest case, i.e. a configuration with instanton number $m=1$.

In that case one decomposes, for example,

\begin{eqnarray}
A_1^{(1)} & = & \sum_{n=1}^{\infty} \sum_{k=0}^{n-1} \ \sqrt{ n
+ 1 - k } | n ; k + 1 >< n + 1 ; k + 1 | + \nonumber \\
& + & \sum_{n=0}^{\infty} | n, n + 1 >< n + 1 , n + 1 | \nonumber
\\
\mapsto & \ & A^{(1)}_1 = \sum_{n_1 = 0 }^{\infty} \sum_{n_2 = 0
}^{\infty} \ \sqrt{ n_1 + 2 } | n_1, n_2 + 1 >< n_1 + 1 , n_2 + 1
| + \nonumber \\
& \ & + \sum_{n_1 = n_2 = 0}^{\infty} | n_1 + 1 , 0 >< 0, n_2 + 1
| . \label{014} \end{eqnarray}

Analogously for the other connection

\begin{eqnarray}
A_2^{(1)} & = & \sum_{n_1=0}^{\infty} \sum_{n_2=0}^{\infty} \
\sqrt{ n_2 + 1 } | n_1, n_2 + 1 >< n_1 , n_2 + 2 |  + \nonumber \\
& + & \sum_{n_1=n_2=0}^{\infty} \sqrt{ 1 + n_2 } | n_1 + 1 , 0 ><
n_1 + 2 , 0 | , \label{015} \end{eqnarray}

Then the quasi-unitary operator looks like:

\begin{eqnarray}
U & = & \sum_{n_1=0}^{\infty} \sum_{n_2 = 0 }^{\infty} | n_1 + 1,
n_2 >< n_1 , n_2 + 1 | + \sum_{n_1=n_2=0}^{\infty} | 0, n_2 >< n_1
+ 1 , 0 | \nonumber \\
U^{\dagger} & = & \sum_{n_1=0}^{\infty} \sum_{n_2 = 0 }^{\infty} |
n_1 , n_2 + 1 >< n_1 + 1 , n_2 | + \sum_{n_1=n_2=0}^{\infty} | n_1
 + 1 , 0 >< 0  , n_2 | \nonumber \\
& \ & . \label{016} \end{eqnarray}

One can check that the commutation rules (\ref{419}) are again
satisfied  since

\begin{eqnarray}
A^{(1)}_1 \overline{A}^{(1)}_1 - \overline{A}^{(1)}_1 A^{(1)}_1 &
= & \sum_{n_1=0}^{\infty} \sum_{n_2 = 0}^{\infty} | n_1, n_2 + 1
>< n_1, n_2 + 1| + \sum_{n_1=0}^{\infty} | n_1 + 1, 0  ><
n_1 + 1, 0 | \nonumber \\
& = & 1 - | 0, 0 >< 0, 0 | = 1 - P_0 = A_2^{(1)}
\overline{A}_2^{(1)} - \overline{A}_2^{(1)} A_2^{(1)} \nonumber \\
A^{(1)}_1 A^{(1)}_2 = A^{(1)}_2 A^{(1)}_1 & = &
\sum_{n_1=0}^{\infty} \sum_{n_2=0}^{\infty} \sqrt{ ( n_1 + 2 )(
n_2 + 1 ) } | n_1, n_2 + 1 >< n_1 + 1 , n_2 + 2 | + \nonumber \\
& + & \sum_{n_1=n_2=0}^{\infty} \sqrt{ 1 + n_2 } | n_1 + 1 , 0 ><
0, n_2 + 2 | \nonumber \\
A^{(1)}_1 \overline{A}^{(1)}_2 = \overline{A}^{(1)}_2 A^{(1)}_1 &
= & \sum_{n_1=0}^{\infty} \sum_{n_2=0}^{\infty} \sqrt{ ( n_1 + 2
)(
n_2 + 1 ) } | n_1, n_2 + 2 >< n_1 + 1 , n_2 + 1 | + \nonumber \\
& + & \sum_{n_1=n_2=0}^{\infty} \sqrt{ 1 + n_2 } | n_1 + 2 , 0 ><
0, n_2 + 1 | . \label{017} \end{eqnarray}

Another possible generalization is extending our solution to
$d=2n$ noncommutative planes. By diagonalizing the $\Theta_{ij}$
matrix as a sum of independent $n$-oscillators, it is not
difficult to generalize our trick. For example let us consider the
six-dimensional case. The basic coordinates commutation relations
are:

\begin{eqnarray}
& \ & [ \hat{x}_1, \hat{x}_2 ] = i \theta_1 \nonumber \\
& \ & [ \hat{x}_3, \hat{x}_4 ] = i \theta_2 \nonumber \\ & \ & [
\hat{x}_5, \hat{x}_6 ] = i \theta_3 . \label{441}
\end{eqnarray}

Introducing three independent raising and lowering oscillator
operators $A_i , i = 1,2,3$ through the relations :

\begin{eqnarray}
X_1 & + & i X_2 = - i \sqrt{\frac{2}{\theta_1}} A_1 \nonumber \\
X_3 & + & i X_4 = - i \sqrt{\frac{2}{\theta_2}} A_2 \nonumber \\
X_5 & + & i X_6 = - i \sqrt{\frac{2}{\theta_3}} A_3 \label{442}
\end{eqnarray}

the six-dimensional instanton is defined through the commutation
relations :

\begin{eqnarray}
& \ & [ A_1, \overline{A}_1 ] = [ A_2, \overline{A}_2 ] = [ A_3,
\overline{A}_3 ] = 1- P_0 \nonumber \\
& \ & [ A_1, A_2 ] = [ A_1, \overline{A}_2 ] = [ A_1, A_3 ] = [
A_1, \overline{A}_3 ] = [ A_2, A_3 ] = [ A_2, \overline{A}_3 ]  =
0 .
\nonumber \\
& \ & \label{443} \end{eqnarray}

By using again the duality between

\beq {\cal H } \times {\cal H } \times {\cal H} \ \ \rightarrow \
\ {\cal H } \eeq

based on the fact that three numbers can be made isomorphic to a
number

\beq ( n_1, n_2, n_3 ) \ \ \rightarrow \ \ n = f( n_1, n_2, n_3 )
\eeq one generalizes the solution (\ref{09}) to six dimensions.

\section{Conclusion}

In this paper we have found a general class of soliton solutions
starting from the Matrix model approach instead of the twistor
approach of Nekrasov and Schwarz. We believe that this intrinsic
characterization of solitons within the matrix model is necessary
to simplify the classification of instantons and to study quantum
mechanics around this nonperturbative solutions. For example the
contribution to the partition function of gauge theory from an
instanton  can be easily derived as an ordinary matrix integral of
the fluctuations around the configuration.

We have started our study from the unitary matrix model, where the
soliton solutions have been characterized as the only solutions of
the matrix model which survive in the $N \rightarrow \infty$ limit
and give a finite contribution to the action.

As a byproduct of this investigation we have reduced a $U(n)$
soliton in terms of a $U(1)$ soliton in two dimensions,
generalizing the well known Morita equivalence of the
noncommutative torus to the noncommutative plane, but we believe
that this can be done in every even dimensions.

Then the solitons within the unitary matrix model have been
compared with the solitons defined in the hermitian matrix model,
finding as characterization the following equation

\beq [ X_i, X_j ] = i \Theta^{-1}_{ij} ( 1- P_0 ) . \label{51}
\eeq

This equation has been analyzed in detail in $d=2n$ dimensions,
and given the general solution, consistent with the equation of
motion

\beq [ X_j, [ X_i, X_j ]] = 0 . \label{52} \eeq

The construction of the soliton solutions in $d=2n$ planes is
based on representing the gauge connection as a sum of two parts,
whose role is to build a projection operator in the commutation
rules (\ref{51}).

We believe that this systematic point of view will help in
classifying all the possible instanton solutions, giving an
accurate and simple scheme of the nonperturbative properties of
the noncommutative gauge theory in $d=2n$.

\end{document}